\documentclass[runningheads]{llncs}
\usepackage[T1]{fontenc}
\usepackage{cite}
\usepackage{graphicx}
\usepackage{multirow}
\usepackage{amsmath,amssymb,amsfonts}
\usepackage{algorithmic}
\usepackage{graphicx}
\usepackage{textcomp}
\usepackage{xcolor}
\usepackage[multiple]{footmisc}
\usepackage{bigfoot}

\DeclareNewFootnote{AAffil}[arabic]
\DeclareNewFootnote{ANote}[fnsymbol]

\begin{document}
\title{MFConvTr: {M}ulti-{F}requency {C}onvolutional {T}ransformer for {F}etal {A}rrhythmia {D}etection in {N}on-{I}nvasive fECG}
%
%
\author{Deva Satya Sriram Chintapenta$^*$ \inst{1}
        \and 
        Aman Verma$^*$ \inst{1}
        \and
        Saikat Majumder\inst{1}}
\authorrunning{D.S.S. Chintapenta et al.}
%
\institute{Department of Electronics and Communications Engineering, \\ National Institute of Technology Raipur, India \inst{1} \\
\email{cdssriram@gmail.com,aman.verma.nitrr@gmail.com,smajumder@etc.nitrr.ac.in}}
\maketitle              
\def\thefootnote{*}\footnotetext{Equal contribution}
\begin{abstract}
NI-fECG have emerged as alternative for fetal arrhythmia monitoring. But due to multi-signal waveform they are tough to understand and due to highly varying and complex nature traditional fiducial methods cannot be applied. Further, it has also been observed that the fetal arrhythmia can be differentiated from the normal signals in both spectral and temporal scales. To this end, we propose Multi-Frequency Convolutional Transformer, a novel deep learning architecture that learns information in contexts with multiple-frequency and can model long-term dependencies. The proposed model utilizes a convolutional-backbone consisting of model Multi-Frequency Convolutions (MF-Conv) and residual connections. MF-Conv in-turn captures multi-frequency contexts in an efficient manner by splitting the input channel and then convoluting each of the splits individually with different kernel size. Accredited to these properties, the proposed model attains state-of-the-art results and that too utilizing very low number of parameters. To evaluate the proposed we also perform extensive ablation studies.

\end{abstract}
\section{Introduction}
Fetal cardiac health monitoring is essential to discover any abnormality in the child before birth. In general, the prevalent procedure to find out cardiac disorder has been arrhythmia detection. 
For adults, Electrocardiogram (ECG) has been a gold-standard for cardiac activity monitoring, hence for arrhythmia prediction in fetus, fECG has been taken in account.
fECG is defined as the electrical physiological signal generated by fetal cardiovascular system \cite{mohebbian2023semi,nakatani2022fetal,rai2023fetal}
Scalp-invasive techniques have been used to measure it during labour, but this limits regular morphological fetal cardiac health monitoring. The method used to monitor fECG should both physically and economically feasible to be repeated as many numbers of times. 
Since, invasive methods follow operation-based strategies they cannot be practiced at a prevalent scale. Non-Invasive fECG (NI-fECG) has emerged as alternative for invasive methods.  Although non-invasive, the NI-fECG signal is amalgamation of Maternal ECG (MECG), the fetal ECG and noise. Due to the same reason, the NI-fECG signals are challenging to comprehend \cite{lin2024advancing}. However, they have merit that they can be	practiced	in	continuous	basis,	with patient	being	at	ease. This can be a key towards fetal healthcare in remote places \cite{zhang2012compressed}. Though NI-fECG is a suitable option but its intricate signal properties hinder proliferated usage. More than being inherently noisy, it comprises of two-separate waveforms. In comprehending the same, there are several other challenges such as its unstructured morphology and incoherent heart-rates of mother and the baby. Hence, in order to robustly model NI-fECG following must be taken in account:

\begin{figure*}[!t]
\small
\centering
\includegraphics[height=7cm,width=1.0\linewidth]{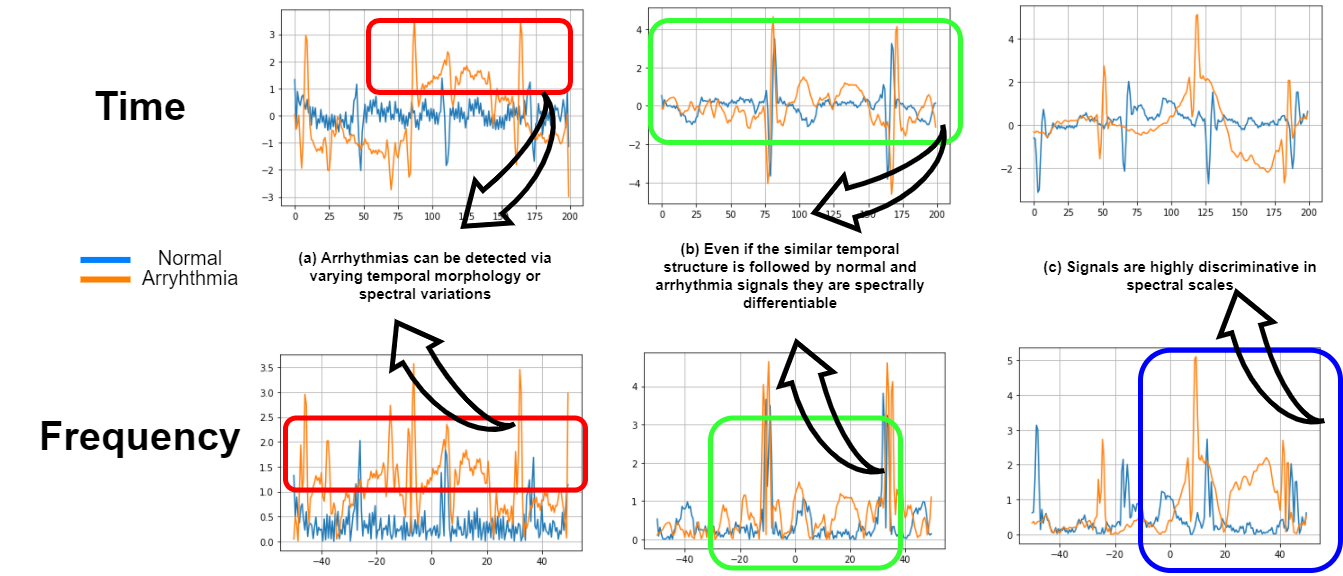}
\caption{Importance of multiple scales for arrhythmia understanding. It is also essential for the model to capture long-term dependencies.} 
\label{fig:FECGCL_Motivation}
\end{figure*}

\begin{itemize}
    \item Arrhythmia occurs in a small-time duration; this implies to the fact that analyzing the signal in multi-frequency scales can be helpful.
    \item It is required to model long-term temporal dependencies in order to formulate linkage in overall temporal-morphology of the input signal.
\end{itemize}

There have been attempts towards capturing contexts within NI-fECG \cite{mohebbian2023semi,rai2023fetal}. However, most of these works formulated frameworks which are non-domain specific. To be specific, \cite{mohebbian2023semi} proposed 1D CNN without consider multi-scale learning within the network. Similarly, \cite{rai2023fetal} utilized general purpose 2D CNNs for NI-fECG analysis.  In order to analyze the importance of different temporal and frequency a visual comparison has been presented in Fig. \ref{fig:FECGCL_Motivation}.  We plotted arrhythmia and normal signal in a single plot for both time and frequency domains. From the figure we conclude that – arrhythmia can be detected by both temporal morphological variations and by abnormalities in spectral band. Even if temporal appearance of the signal is not discriminative, in frequency domain – normal and arrhythmia signals are differentiable. Further, arrhythmia might exist for short time span but they do affect the entire waveform for a longer-duration.NI-fECG involves both high-frequency and low-frequency signals, arrhythmia themselves occur in a short span of time locally while the effect remains present at global scales. Thus, multi-scale and multi-frequency capture can be instrumental in discriminating between arrhythmia and normal signal. To leverage these motivations, this research proposes MFConvTr - a deep-learning architecture to capture long-term dependencies at multi-scale level for robust classification of fetal arrhythmia in NI-fECG signal. Following are the key contributions:

\begin{itemize}
    \item We propose a novel deep-learning model MFConvTr to capture multi- frequency contexts and structure long-term temporal dependencies for robust classification of arrhythmia.
    \item MFConvNet is proposed which learns features at multiple scales in a parametrically efficient manner. It leverages domain-specific and residual learning techniques to formulate backbone of MFConvTr. Its parametrically efficient design lets it operate at edge.
    \item Transformer\cite{vaswani2017attention} network is utilized which employs multi-head self-attention mechanism to understand long-term dependencies.
    \item Extensive experimentation and ablation study over benchmark NI-fECG Arrhythmia database is performed. State-of-the-art results are obtained for the same.
\end{itemize}

The rest of this paper is organized as follows. Related previous work is reviewed in Section 2. Then, in Section 3, we discuss the proposed MFConvTr. In Section 4, we discuss the experimental protocol and obtained results. We finally conclude the paper in Conclusion section.

\section{Related works}
As mentioned earlier, it is challenging to extract discriminative features from the NI-fECG signal due to inherent noise and multi-signal waveform. Hence, it has been a key research question – ‘How to separate fECG from MECG?’. To this end, research in \cite{mohebbian2021fetal} utilized Attention-Mechanism along with CycleGAN to achieve the state-of-the-art results. A shortcoming in the former has been the metrics used can create-false picture of replication. Further, there is a requirement of domain-specific models along with evaluation strategies such as open-set cross validation, which can infer out actual performance in unconstrained scenarios.

Predicting directly on NI-fECG can be an alternative, if robust models are designed. 
Some of the initial attempts \cite{al2022machine,keenan2022detection} involved machine learning based classifiers. But these methods relied upon fiducial features which can vary significantly across subjects, sessions and acquisition devices.  Utilizing automatic feature extraction techniques such as deep learning, then became major research direction \cite{rai2023fetal,nakatani2022fetal}. To be specific, \cite{ganguly2020non} firstly proposed usage of 1D CNNs. But the same work proposed usage of vanilla 1D CNNs, and evaluated performance on less challenging protocols. Similarity, \cite{nakatani2022fetal} and \cite{mohebbian2023semi} utilized pooling based 1D CNN. This method similar to previous method \cite{ganguly2020non} did not any signal specific modeling mechanism. In \cite{rai2023fetal} authors utilized generic 2D CNNs which again questions the generalizability of the proposed method. 
Most of the methods either employed non-domain specific design or did not consider discriminability at multiple scales. Furthermore, long-range dependencies have also not been considered. To this end, we propose MFConvTr which ensures robust modeling along with low parametric design. 

\section{Proposed methodology and MFConvTr}

\begin{figure*}[!t]
\small
\centering
\includegraphics[height=6cm,width=1.0\linewidth]{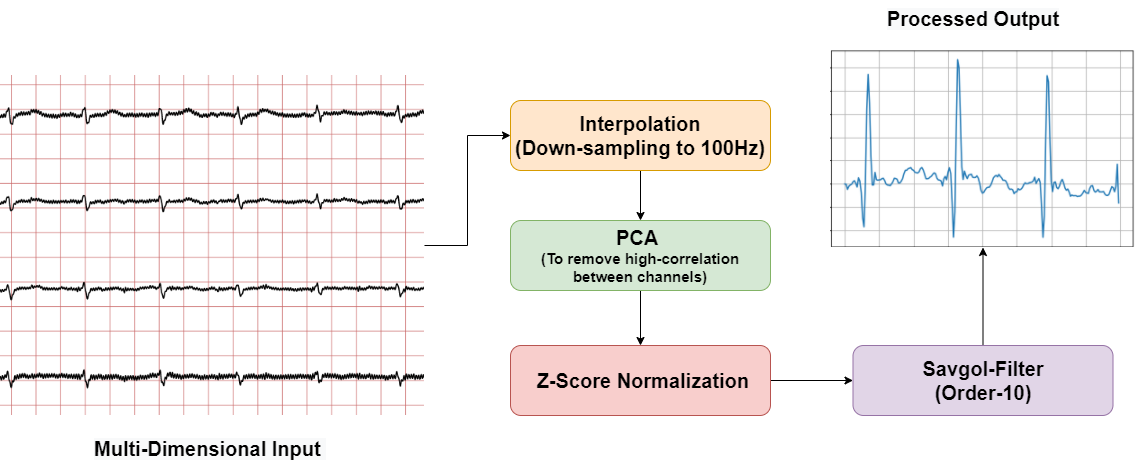}
\caption{The proposed signal preprocessing method for MFConvTr. This involves channel reduction using PCA and noise filtering using Savgol filter.} 
\label{fig:FECGCL_preprocess}
\end{figure*}

\begin{figure*}[!t]
\small
\centering
\includegraphics[height=8cm,width=12cm]{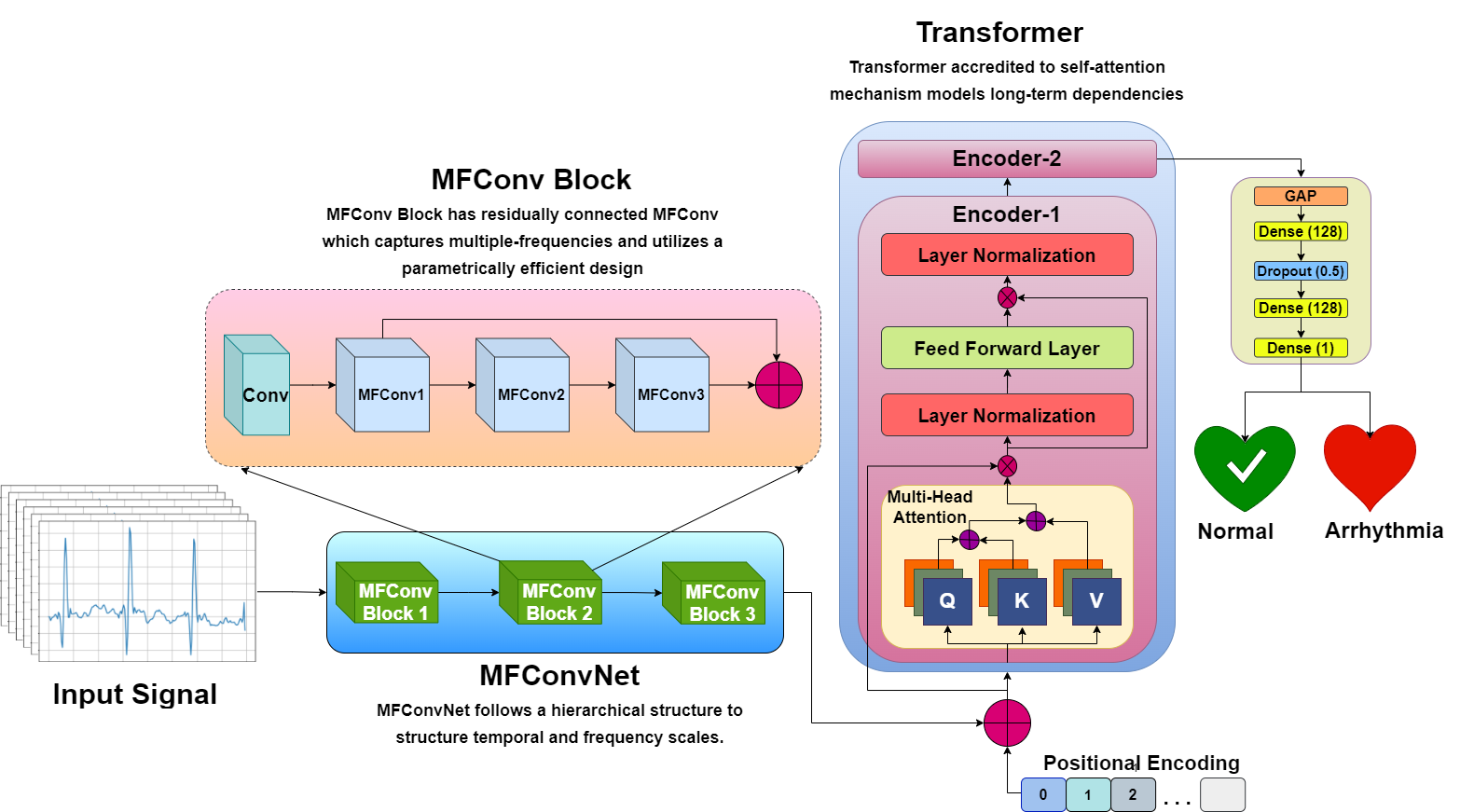}
\caption{The proposed MFConvTr. Firstly, MFConvNet extracts multi-scale features, and then long-range dependecies are modeled using transformer network.} 
\label{fig:MFConvTr}
\end{figure*}

\subsection{Signal preprocessing}
We present a novel pipeline to preprocessing the inputs for MFConvTr (illustrated in Fig. \ref{fig:FECGCL_preprocess}). Since, the input NI-fECG signal is a multi-channel signal, we take the first 4 channels as the input (some records had a channel extra, there we took the first 5 channels only). Then using interpolation we down sample the input record to 100Hz. The input channels in a record are highly correlated, thus we utilized PCA to formulate a new-basis. Therefore, a single channel is extracted using PCA. We then perform Z-Score normalization. Finally, a Savitzky-Golay Filter of order 10 and window-length 17 is applied for smoothing the signal. In this work, we utilize signal length of 2 seconds. as well as 100 Hz sampling frequency. This simulates practical scenarios wherein acquisition devices may not be advanced. Also, this aspect has not been given much importance in the existing literature \cite{lin2024advancing,ganguly2020non,rai2023fetal,mohebbian2023semi}.

\subsection{MFConvTr: Overview}
This section presents the overall deep-learning architecture of the MFConvTr. As shown in Fig. \ref{fig:MFConvTr}, MFConvTr is composed of two major component - (i) MFConvNet and (ii) Transformer. Each of the components has their individual responsibility, MFConvNet works as the backbone of the MFConvTr and extracts features with multi-frequency contexts. While, Transformer module learns temporal features and captures long-term structure present in the input NI-fECG signals. The MFConvNet follows a hierarchical architecture with residual connections in between, this lets us preserve the original spatial contexts and capture rich-features in temporal scales. After the Transformer model, Global Average Pooling is applied. Finally, after a stack of fully connected layers final classification is performed. The overall architecture utilizes limited number of parameters and is robust for detecting arrhythmia. Each of the components of MFConvTr are explained in following subsections.

\begin{figure}[!t]
\small
\centering
\includegraphics[width=1.0\linewidth]{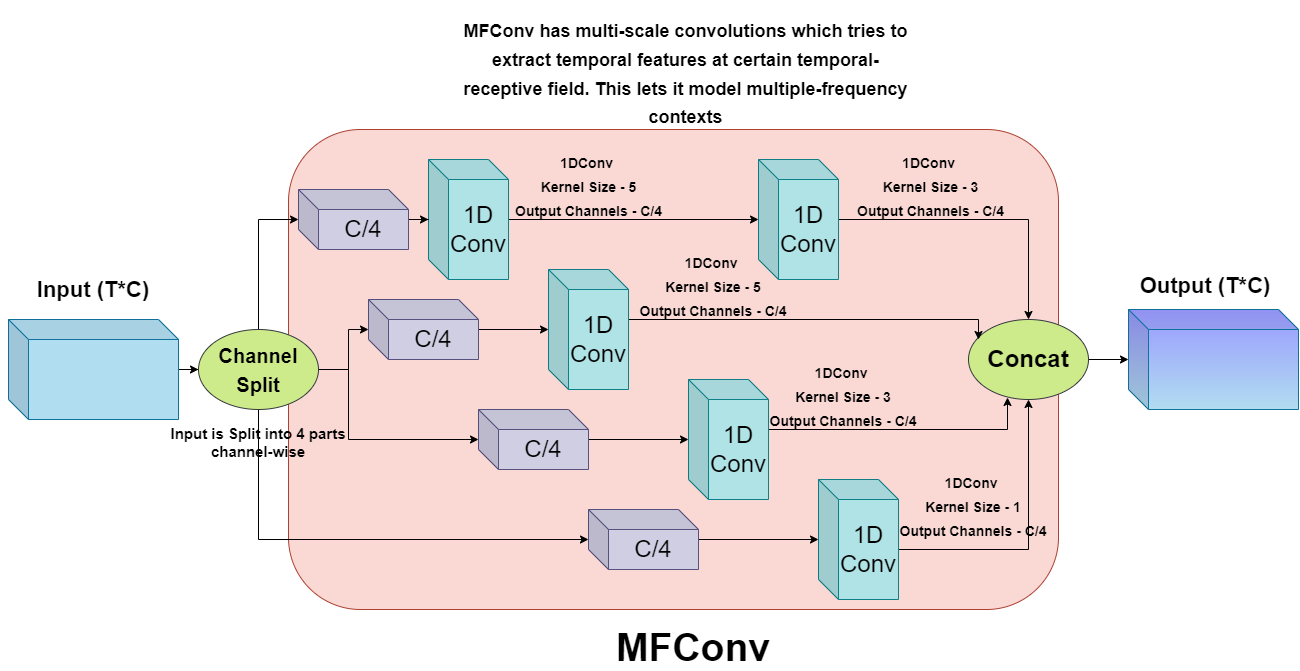}
\caption{The proposed MFConv. Accredited to multiple scale in temporal processing, it can capture different spectral contexts as well.} 
\label{fig:MFConv}
\end{figure}

\subsection{MFConvNet}
In this subsection, we discuss MFConvNet, the backbone of MFConvTr. As illustrated in Fig. \ref{fig:MFConvTr}, MFConvTr is composed of $3$ MFConv blocks. At the begining of every MFConv block, there is 1D convolution present which increases channel dimensions while correlating the features extracted at different frequency scales. The MFConv blocks output 32, 64 and 128 channels respectively. Accredited to limited channels at the input of every convolution, the number of parameters in the model is highly optimized. Addition of Residual connections helps in stabilizing training. We did not pool in the backbone network so as to preserve temporal structure for the transformer network. While the residual connections in architecture also contribute in the same. MFConvNet utilizes the proposed Multi-Frequency Convolutions (MFConv). The working diagram of MFConv has been illustrated in Fig. \ref{fig:MFConv}. MFConv first splits the input into 4 parts channel-wise. Then, each of the part is individually convoluted with kernels of different sizes, we choose the sizes to be 15, 5, 3 and 1. Since, the kernel sizes are varying, MFConv views the information at multiple level and similar to DWT. Hence, the overall architecture of MFConv operates to enforce learning at multi-frequency contexts. We further split the kernel of size (15) into sub-kernels of 3 and 5. This all helps in reducing the parameters. Each of the parts are operated through 1D-Convolutions performed at different scales.

\subsection{Transformer}
Transformer network is instrumental in capturing long-term dependencies. To be specific, in MFConvTr we utilize $2$ encoder heads. Each encoder consists of multi-head attention block, layer normalizations and Feed- Forward layer. Self-Attention is applied via multi-head attention block, which first projects each time-step into certain dimension d : Query-Q, Key- K, Value-V. Then, it applies SoftMax based attention over each of the steps so as to model correlation amongst the different temporal-steps. Multiple attention heads are used to capture different understanding at each head. On overall basis, the self-attention mechanism helps in capturing relevant information on temporal scales. 

\section{Experimental analysis}

\subsection{Dataset used and experimental protocol}
We perform all the experimentation over NIFEA-DB \cite{goldberger2000physiobank}. It consists of data from 26 records with 14 being normal while the other 12 having Arrhythmia. Each record had recording of about 7-32 minutes and number of electrodes varied from 4 to 5. Sampling rate was either 500 Hz or 1000 Hz. For conducting ablation studies and comparison we consider 50-50\% division of a record on the basis of time. First 50\% of all the records was used in training while the remaining in testing. We consider average accuracy as the performance metric. 

\subsection{Comparative study with State-of-the-art methods}
In Table \ref{tab:Comp}, the results of comparative study of the proposed model and its variants (ConvNet, Tx, MFConvNet, ConvNet+Tx, MFConvNet+BiLSTM). The proposed model attains highest performance of 93.67\% while using 2x less parameters than the second-best model. It is clear from the same, that the proposed model is performing well in comparison as . The proposed MFConv attains the highest performance with high-parameter efficiency. With addition of transformer there is almost 1\% increment in performance, same is the case when the proposed model is compared with MFConvNet+BiLSTM. Also, the proposed model outperforms most of the state-of-the-art methods \cite{al2022machine,nakatani2022fetal,rai2023fetal,mohebbian2023semi}. These gain in performance in comparison to 1D CNNs \cite{nakatani2022fetal,mohebbian2021fetal} can be accredited to multi-scale feature learning and modeling long-term dependencies. It must also be noted that the number of parameters is also very less, and hence is optimal edge use cases. 

\begin{table}[]
\centering
\caption{Comparative study with State-of-the-art methods and variants of MFConvTr. Transformer is mentioned as 'Tx'} \label{tab:Comp}
\begin{tabular}{|c|c|c|c|}
\hline
\textbf{Method}                        & \textbf{Data Split} & \textbf{Acc. (\%)} & \textbf{Parameters} \\ \hline
\textbf{ConvNet}                       & 50:50               & 93.58              & 1155804             \\ \hline
\textbf{Tx}                            & 50:50               & 92.19              & 452417              \\ \hline
\textbf{MFConvNet}                     & 50:50               & 92.73              & 256841              \\ \hline
\textbf{ConvNet+Tx}                    & 50:50               & 93.00              & 1420769             \\ \hline
\textbf{MFConvNet+BiLSTM}              & 50:50               & 92.48              & 52009               \\ \hline
\textbf{Decision Tree \cite{al2022machine}} & -                   & 93.12              & -                   \\ \hline
\textbf{Semi-Supervised 1D CNN \cite{mohebbian2023semi}}        & -                   & 92.00              & -                   \\ \hline
\textbf{Deep CNN \cite{sharma2021deep}}         & 85:15               & 96.00              & -                   \\ \hline
\textbf{Deep QRS Complex \cite{zhong2018deep}}              & 85:15               & 91.33              & -                   \\ \hline
\textbf{1D CNN \cite{nakatani2022fetal}}            & -                   & 95.00              & -                   \\ \hline
\textbf{Entropy \cite{keenan2022detection}}            & -                   & 90.00              & -                   \\ \hline
\textbf{ResNet152 \cite{rai2023fetal}}         & -                   & 86.41              & 25.6 million        \\ \hline
\textbf{Xception \cite{rai2023fetal}}          & -                   & 91.34              & 22.9 million        \\ \hline
\textbf{Wavelet 1D CNN \cite{ganguly2020non}}  & 50:50               & 96.00              & -                   \\ \hline
\textbf{MFConvTr (Proposed)}           & 50:50               & \textbf{93.67}     & \textbf{521801}     \\ \hline
\textbf{MFConvTr (Proposed)}           & 85:15               & \textbf{96.32}     & \textbf{521801}     \\ \hline
\end{tabular}
\end{table}

\subsection{Ablation study on MFConvNet}

\begin{table}[!t]
\caption{Ablation Study on MFConvNet: results for different configurations splits and kernel sizes. These are reported for MFConvNet and MFConvTr. We have highlighted the results for the proposed. These experiments are conducted for 50:50 split.}
\centering
\label{tab:MFConvNet}
\begin{tabular}{|c|c|cc|cc|}
\hline
\multirow{2}{*}{\textbf{\# Splits}} & \multirow{2}{*}{\textbf{Kernel size}} & \multicolumn{2}{c|}{\textbf{MFConvNet}}                     & \multicolumn{2}{c|}{\textbf{MFConvTr}}                      \\ \cline{3-6} 
                                    &                                       & \multicolumn{1}{c|}{\textbf{Acc. (\%)}} & \textbf{Paramters} & \multicolumn{1}{c|}{\textbf{Acc. (\%)}} & \textbf{Parameters} \\ \hline
\multirow{4}{*}{1}                  & 1                                     & \multicolumn{1}{c|}{74.03}             & 109833             & \multicolumn{1}{c|}{66.26}            & 373793              \\ \cline{2-6} 
                                    & 3                                     & \multicolumn{1}{c|}{91.47}             & 258401             & \multicolumn{1}{c|}{91.29}            & 523361              \\ \cline{2-6} 
                                    & 5                                     & \multicolumn{1}{c|}{91.29}             & 407969             & \multicolumn{1}{c|}{91.94}            & 672929              \\ \cline{2-6} 
                                    & 15                                    & \multicolumn{1}{c|}{93.58}             & 1155809            & \multicolumn{1}{c|}{93.00}            & 1420769             \\ \hline
\multirow{3}{*}{2}                  & 15,1                                  & \multicolumn{1}{c|}{93.60}             & 333617             & \multicolumn{1}{c|}{91.78}            & 598877              \\ \cline{2-6} 
                                    & 15,3                                  & \multicolumn{1}{c|}{93.22}             & 365873             & \multicolumn{1}{c|}{92.68}            & 630833              \\ \cline{2-6} 
                                    & 15,5                                  & \multicolumn{1}{c|}{93.13}             & 398129             & \multicolumn{1}{c|}{92.57}            & 663089              \\ \hline
\multirow{3}{*}{3}                  & 15,1,3                                & \multicolumn{1}{c|}{92.68}             & 273659             & \multicolumn{1}{c|}{93.38}            & 538619              \\ \cline{2-6} 
                                    & 15,3,5                                & \multicolumn{1}{c|}{92.27}             & 302873             & \multicolumn{1}{c|}{93.11}            & 567833              \\ \cline{2-6} 
                                    & 15,5,1                                & \multicolumn{1}{c|}{92.61}             & 289043             & \multicolumn{1}{c|}{92.70}            & 559003              \\ \hline
\textbf{4}               & 15,1,3,5                              & \multicolumn{1}{c|}{92.73}             & \textbf{256841}    & \multicolumn{1}{c|}{\textbf{93.67}}   & \textbf{521801}     \\ \hline
\end{tabular}
\end{table}

In order to understand significance of multi-frequency convolutions, we perform ablation study on MFConv. In this experimentation, at each of the model we modify the number of splits and also the kernel sizes. This allowed us to get insights over the importance of multiple-frequency contexts. Higher Kernel size represents capturing low-frequency details while the converse is also true. The results obtained in this experiment are tabulated in Table \ref{tab:MFConvNet}. It is evident from these results that low-frequency information has the higher prominence but adding multi-frequency contextual information has two-fold benefits. Firstly, it helps model extract richer features and secondly, it contributes in reduction of parameters to a great extent. Further, when multi-context information is passed to transformer it brings performance gain. This is because the attention-heads of the transformer model now has access to multi-contextual representations in its input. The proposed model achieves highest performance while conserves 2x parameters in comparison to high-end performing MFConvNet model with number of splits 1 and kernel size as 15.

\subsection{Ablation study on Transformer}
Finally, to leverage full-potential in the transformer model we experimented over the number of attention heads and encoder layers. We kept the MFConvNet model fixed while the hyperparameters of the transformer model were changed. The results obtained in this ablation have been tabulated in Table \ref{tab:Tx}. It is clear from the following table, that when the number of attention-heads are 8 while encoder layers are set to 2, we obtain the optimal results of 93.67\%. When the number of multi-heads in self- attention block are higher than required, correlated information flows through and result is fall in performance. Further, this reasoning also follows for increase in encoder layers.

\subsection{Error analysis}

\begin{figure}[!t]
\small
\includegraphics[width=1.0\linewidth]{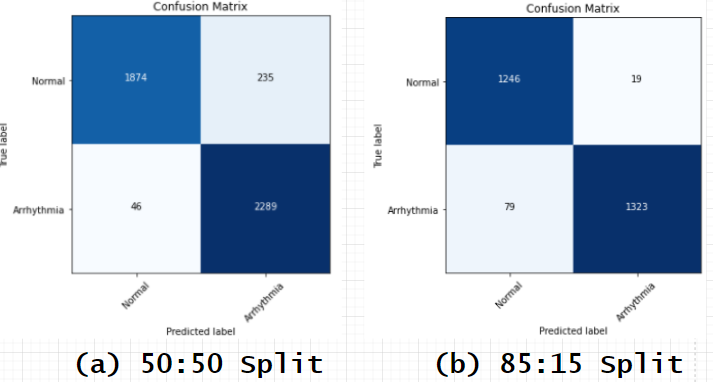}
\caption{Confusion matrix for MFConvTr on the different settings.} 
\label{fig:CM}
\end{figure}

\begin{table}[!t]
\caption{Ablation Study on Transformer: Results for different number of encoders and attention heads.}
\centering
\label{tab:Tx}
\centering
\begin{tabular}{|c|c|c|c|}
        \hline
        \textbf{$\#$ Encoders} & \textbf{$\#$ Attention Heads} & \textbf{Acc. (\%)} & \textbf{Parameters} \\
        \hline
        1 & 4 & 93.04 & 389322 \\
        \hline
        1 & 8 & 92.79 & 389322 \\
        \hline
        1 & 16 & 92.21 & 389322 \\
        \hline
        2 & 4 & 93.02 & 521801 \\
        \hline
        2 & 8 (Proposed) & \textbf{93.67} & 521801 \\
        \hline
        2 & 16 & 93.36 & 521801 \\
        \hline
        3 & 4 & 92.64 & 654281 \\
        \hline
        3 & 8 & 93.51 & 654281 \\
        \hline
        3 & 16 & 92.12 & 654281 \\
        \hline
    \end{tabular}
\end{table}

We perform error analysis in this subsection. To this end, we plot confusion matrix in Fig. \ref{fig:CM}. It is clear from these that Intrinsic bias is towards classifying normal signals as arrhythmia. The possible reason for this is noisy-temporal structure of the normal signal. Also, we find that there are more errors when 50:50 data split is used. This could be because transformer requires large number of examples for training.

\section{Conclusion}
This research introduced novel MFConvTr model which learns features at multiple-frequency contexts and can understand long- term dependencies for robust detection of arrhythmia. MFConvTr obtained state-of-the-art results, that too utilizing very low number of parameters.
We signified the importance of multi-frequency contexts while we introduced novel and parameter efficient MFConv to capture multi-frequency information while remaining in temporal-domain. In future work, we will be focusing on multi- branch 1D convolutional network and upon octave convolutions. We shall further be exploring performance of the model over different time periods and sampling rates.

\bibliographystyle{splncs04}
\bibliography{main}{}
\end{document}